\documentclass[aps,prl,reprint,twocolumn,superscriptaddress,longbibliography,nofootinbib,preprintnumbers,floatfix]{revtex4-1}
\usepackage{amsmath,amssymb}
\usepackage{times}
\usepackage{graphicx}
\usepackage{natbib}
\usepackage{hyperref}
\usepackage{bm}
\usepackage{slashed}
\usepackage{xcolor}
\usepackage{verbatim}
\usepackage{MnSymbol}

\newcommand{\bw}{\begin{widetext}}
\newcommand{\ew}{\end{widetext}}
\newcommand{\be}{\begin{equation}}
\newcommand{\ee}{\end{equation}}
\newcommand{\bestar}{\begin{equation*}}
\newcommand{\eestar}{\end{equation*}}

\newcommand{\bi}{\begin{itemize}}
\newcommand{\ei}{\end{itemize}}
\newcommand{\bea}{\begin{eqnarray}}
\newcommand{\eea}{\end{eqnarray}}

\newcommand{\hbo}{\hbox to 1 true cm {\hfill } }

\newcommand{\vc}[1]{\mbox{\boldmath$#1$}}

\newcommand{\ud}{\mathrm{d}}

\newcommand{\E}{\boldsymbol{E}}
\newcommand{\B}{\boldsymbol{B}}

\newcommand{\azero}{a_0}
\newcommand{\apeak}{\hat{a}_0}
\begin{document}

\title{The intensity dependent mass shift: existence, universality and detection}

\author{Chris Harvey}
\email{christopher.harvey@physics.umu.se}
\affiliation{Department of Physics, Ume\aa\ University, SE-901 87 Ume\aa, Sweden}

\author{Thomas Heinzl}
\email{theinzl@plymouth.ac.uk}
\affiliation{School of Computing and Mathematics, University of Plymouth, Plymouth PL4 8AA, UK}

\author{Anton Ilderton}
\email{anton.ilderton@physics.umu.se}
\affiliation{Department of Physics, Ume\aa\ University, SE-901 87 Ume\aa, Sweden}

\author{Mattias Marklund}
\email{mattias.marklund@physics.umu.se}
\affiliation{Department of Physics, Ume\aa\ University, SE-901 87 Ume\aa, Sweden}

\begin{abstract}

The electron mass shift in a laser field has long remained an elusive concept. We show that the mass shift can exist in pulses but that it is neither unique nor universal: it can be reduced by pulse shaping. We show also that {the detection of mass shift effects in laser-particle scattering experiments} is feasible with current technology, even allowing for the transverse structure of realistic beams.

\end{abstract}


\maketitle
{The frequency shift of radiation emitted by a particle passing through an oscillatory electromagnetic field can be attributed to an effective increase in the particle's mass \cite{Kibble:1965zz,Eberly:1968,Dodin}. This effect is seen daily in undulators, in which electrons pass through a spatially varying magnetic field \cite{Review1}, providing the basis of XFELs \cite{Elias}. An unambiguous signature of the analogous mass shift in a laser pulse has so far not been obtained~\cite{Mowat:1971ry,McD-talk}.

In both cases, the theory behind the mass shift is based on the assumption of periodic, essentially univariate fields \cite{Lau:2003,COMP}. This is a good description of the regular, well understood magnetic fields of an undulator \cite{DAVV}. For laser fields, the mass shift was originally described using a monochromatic plane wave model \cite{Sengupta,Volkov},  and one reason for the lack of an observation in this case is transverse size effects. These are known to overwhelm mass shift signals at low intensity \cite{50691} and will also be important at high intensities, which are obtained by tightly focussing the laser. Nevertheless, multi-photon effects predicted by plane wave models have been observed in a moderate intensity regime, including higher harmonic generation \cite{Chen:1998} and nonlinearly scattered electron yields \cite{Bamber:1999zt}.}

{Despite both classical and quantum theories permitting an exact treatment of plane wave background fields, the mass shift has also remained theoretically elusive.} The ``lore'' of strong-field QED is based on \cite{Reiss1, Nikishov:1963} which almost exclusively dealt with the idealised case of monochromatic waves (zero bandwidth) and which therefore pushed the mass shift to the forefront. {However, when one considers scattering in pulses (nonzero bandwidth) \cite{68016,Boca:2009zz,Seipt:2010ya,Mackenroth:2010jr,Ilderton:2010wr,Seipt:2011dx,Heinzl:2010vg}, there has been confusion over whether the mass mass shift of the monochromatic case, and its effects, are always present or not \cite{debate}.} The following questions therefore remain unsettled. What are the circumstances leading to a mass shift? When a mass shift emerges, how universal is it? {Is it possible to observe mass shift effects in laser-particle scattering?} 

We answer these questions below.  We first review the emergence of the mass shift and its impact on photon emission spectra. We then construct examples of fields which yield a lower mass shift due to pulse shaping \cite{Haynam}. This new mass shift is shown explicitly to control the emission spectra. {Finally, we identify the moderate intensity regime in which transverse size and short pulse effects can be counterbalanced, allowing a measurement of mass shift effects in laser-particle scattering}. This regime is already accessible to experiment.

\paragraph{Conventions.}
A plane wave travelling in the negative $z$-direction is characterised by a null wave vector $k_\mu = \omega (1,0,0,1)$,  with central frequency $\omega$.  The field strength of the wave can depend arbitrarily on $\phi:=k.x$.  Experimentally, one begins with a finite amount of energy which can be formed, at least in principle, into different pulse shapes. It is therefore energy which should be fixed in order to study the effects of pulse-shaping. The energy in a plane wave (per unit transverse area) is proportional to $\mathcal{E} := {\int\!\ud  \phi}\ \E^2(\phi)$ since $\E^2=\B^2$. In light of this, a useful parameter is $\azero$, the root-mean-square (r.m.s.) (rather than peak) intensity of the pulse \cite{A0,Heinzl:2008rh}. For an $N$-cycle pulse (duration $2\pi N$ in $\phi$), $a^2_0$ is
\be\label{AZERO}
	\azero^2 := \frac{e^2}{m^2\omega^2} \frac{\mathcal{E}}{2\pi N}  \equiv \frac{e^2E_\text{rms}^2}{m^2\omega^2} \;,
\ee
and for periodic fields, $\azero$ coincides with the cycle r.m.s. Our plane wave may then be written (we take linear polarisation from here on) $F_{\mu\nu}(\phi) = (a_0 m /e) f'(\phi) (k_\mu l_\nu - l_\mu k_\nu)$,  with polarisation vector $l_\mu = \delta^1_\mu$. The profile function $f'$ must be normalised so that its r.m.s.\ over the pulse is unity, in order to respect (\ref{AZERO}).  {Let the pulse turn on at $\phi=0$; we choose the gauge potential $eA_\mu(\phi) = a_0 m f(\phi) l_\mu$ with $f=0$ for $\phi\leq 0$.}

\paragraph{{Monochromatic plane waves.}}
{Consider photon emission by an electron, $e (p) \to e(p') + \gamma(k')$, in a monochromatic, and therefore periodic, plane wave. One finds that emission rates are built from a sum over sub-processes governed by the conservation of  \emph{quasi}-momentum, $q_\mu := p_\mu + (\azero^2 m^2/2 k.p) k_\mu$ ($p\to p'$ for the outgoing $e^-$) according to \cite{Nikishov:1963}}
\be\label{1bad}
	q_\mu + n k_\mu = q'_\mu + k'_\mu  \;, \quad n = 1, 2, 3 \ldots \; ,
\ee
with the appearance of $n$ laser photons. Squaring the quasi-momentum yields the shifted mass squared 
\be \label{MASSSHIFT}
  q^2 = m_*^2 := m^2 (1 + \azero^2) \; .
\ee
Being averaged quantities, neither quasi-momenta nor the mass shift can be observed directly. However, their effects can be seen in the photon emission spectra. Due to the perfect periodicity of the monochromatic wave, the spectral density is an (unphysical) delta comb of infinitely sharp and strong peaks supported on a discrete set of frequencies determined by (\ref{1bad}): this is made explicit by squaring and rearranging (\ref{1bad}) to obtain a modified Compton formula for the frequencies $\omega_n^\prime$ in terms of the asymptotic momenta $p$ and $p'$. A particularly simple example, which makes the mass shift's role clear, is obtained for initial conditions such that ${\bf q}=0$, for then the spectral peak positions/frequencies $\omega_n^\prime$ become \cite{LL}
\be
\label{w3}
\begin{split}
	\frac{1}{\omega_n^\prime} &= \frac{1}{n\omega} + \frac{1}{m_*}(1-\cos\theta)  \; ,
\end{split}
\ee
with $\theta$ the scattering angle between $\vc{k}$ and $\vc{k}'$: this is the standard Compton formula with $m\to m_*$ and an incoming photon frequency $n\omega$. In this case, the Compton red shift is reduced as the laser photons transfer less energy to the heavier electron. In general, the lab frame spectra depend sensitively on the relative strengths of $a_0$ and electron $\gamma$, see \cite{Harvey:2009ry,Heinzl:2009nd}.
%
%
%%%%%%%%%%%%%%
\paragraph{Flat-top pulses.}
%%%%%%%%%%%%%%
%
%
Monochromatic waves are crude models of realistic laser beams, which are unavoidably pulsed, i.e.\ of finite duration in $\phi$.  {(For undulators, finite pulse duration is analogous to the unavoidably finite spatial extent of the undulator itself.)} This may be described by modulating the monochromatic fields by an envelope of finite width $2\pi N$ for a pulse of $N$ cycles. Only mild modifications are expected, compared to the monochromatic case, when $N \gg 1$ (i.e.\ for limited bandwidth \cite{Kibble:1965zz}), but if $N$ is small (a few cycles only) then the pulse may become strongly distorted.

We begin here with the simplest pulse, given by a finite wave train \cite{68016}, meaning a flat top (rectangular) envelope. Such waves are ``almost periodic" in that the single cycle pattern is exactly repeated $N$ times. Let us consider how finite pulse duration affects the photon emission spectra. While for monochromatic waves there was only a single frequency $\omega$, and we obtained a line spectrum, a flat-top pulse has a finite bandwidth and therefore a (small) range of frequencies. The delta-comb spectrum is replaced by a diffraction pattern exhibiting broadened peaks and side bands \cite{Krafft:2003is,Heinzl:2009nd}. In order to obtain quantitative results for the photon emission rates in this (slightly) more realistic scenario one first notes that momentum conservation is expressed in terms of ordinary asymptotic momenta (as it is for more general pulses) \cite{Mackenroth:2010jr,Ilderton:2010wr,Hartin:2011vr},
\be\label{1}
	p_\mu + s k_\mu = k'_\mu + p'_\mu \; , \quad s > 0 \;,
\ee
where $s$, a longitudinal momentum fraction Fourier conjugate to $\phi$, parameterises the continuous frequency spectrum of the laser, equivalently the range of energies absorbed in the scattering process. One finds, though, that the emission peaks remain located, for given $\azero$, at precisely the frequencies $\omega_n'$ which follow from the \emph{quasi-}momentum conservation law (\ref{1bad}). How can this be reconciled with (\ref{1})? {Here one must consider not only momentum conservation but also the details of the emission rates.} It emerges that, because the field is (almost) periodic, the peaks in the diffraction pattern correspond to  constructive interference and are associated to particular $s$ values which we call $s_n$. These give scattered photon frequencies $\omega'(s_n)$.  The $s_n$ are determined by the dynamics and may be calculated exactly in our simple field, as in \cite{Heinzl:2010vg}, and are
\be\label{S-VALUES}
	s_n \equiv n + \frac{\azero^2 m^2}{2k.p} - \frac{\azero^2 m^2}{2k.p'} \;, \quad n = 1,2,3, \ldots \;.
\ee
Inserting this into (\ref{1}), one precisely recovers (\ref{1bad}), with the same mass shift (\ref{MASSSHIFT}) and thus $\omega'(s_n) \equiv \omega'_n$ found from (\ref{1bad}): being controlled by the same shifted mass the peak positions $\omega' =  \omega_n^\prime$ in the radiation spectra for monochromatic waves and flat top pulses \emph{must} coincide. The observation of the peaks, at the predicted frequencies and angles, would confirm the presence of the mass shift, even in a pulse.

{In general, the emission rates will tell us whether a mass shift is present or not. These rates are built from the semiclassical Volkov wavefunctions which describe exactly the influence of the background on the fermions \cite{Volkov}; the current carried by these wavefunctions is that of a particle following the classical orbit determined by the Lorentz force in a plane wave \cite{Nikishov:1963,LL}. Extracting information on the mass shift from the rates is, in general, difficult, but progress can be made when we retain periodicity of the fields (including fields which are `periodic for a finite duration' such as the flat-top pulses above).  In this case, identifying the support of the strong peaks, i.e., finding (\ref{S-VALUES}), corresponds to identifying the quasi-momentum with the {\it average classical momentum over a laser cycle} \cite{Heinzl:2010vg}. This defines the quasi-momentum, and thus the mass shift, in a periodic field. We therefore give the averaged classical momentum explicitly. Let a particle of initial momentum $p_\mu$ enter a plane wave at $\phi=0$. The solution of the Lorentz force equation is a textbook result, see \cite{HARTEMANN}. Writing the cycle average as $\langle\ \rangle$ , the quasi-momentum is then
\be\label{KINETIC}
	q_\mu := p_\mu - e \langle A\rangle +\frac{2ep.\langle A\rangle-e^2\langle A^2\rangle}{2k.p}k_\mu \;,
\ee
for $A_\mu(\phi)$ as above. This will be useful below. Taking $f(\phi)=\sin(\phi)$ one recovers the monochromatic results (\ref{1bad}) and (\ref{MASSSHIFT}).}
%
%%%%%%%%%%%%%%%%
\paragraph{Mass shift reduction.}
%%%%%%%%%%%%%%%%
%
\begin{figure}[t!]
\includegraphics[width=0.55\columnwidth]{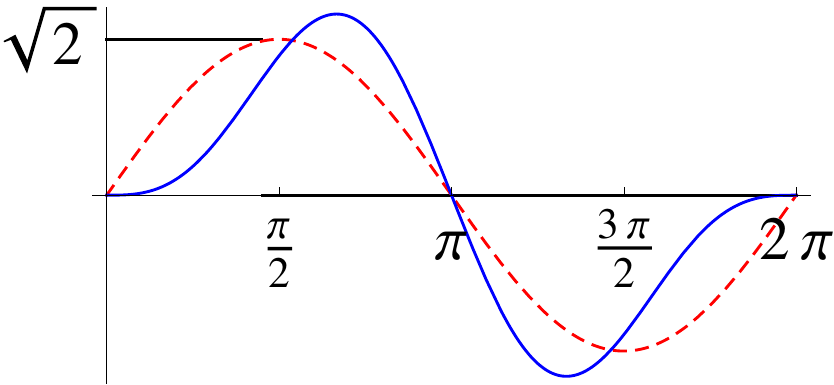}
\caption{\label{FIG:ONECYCLE} The profiles (\ref{PulseZero}) [red/dashed] and (\ref{PulseOne}) [blue/solid], both normalised to contain the same energy.}
\end{figure}
Having addressed the cases for which the standard mass shift (\ref{MASSSHIFT}) emerges, together with its spectral consequences, let us now compare different pulse shapes. This will allow us to assess the universality (or otherwise) of the mass shift. The existence of a mass shift in general will be touched on below; here we will show explicitly that there are cases in which nonstandard mass shifts emerge (without changing $a_0$) leading to distinct signals in the emission spectrum. This is achieved by pulse shaping.
\begin{figure}[t!!]
\includegraphics[width=0.85\columnwidth]{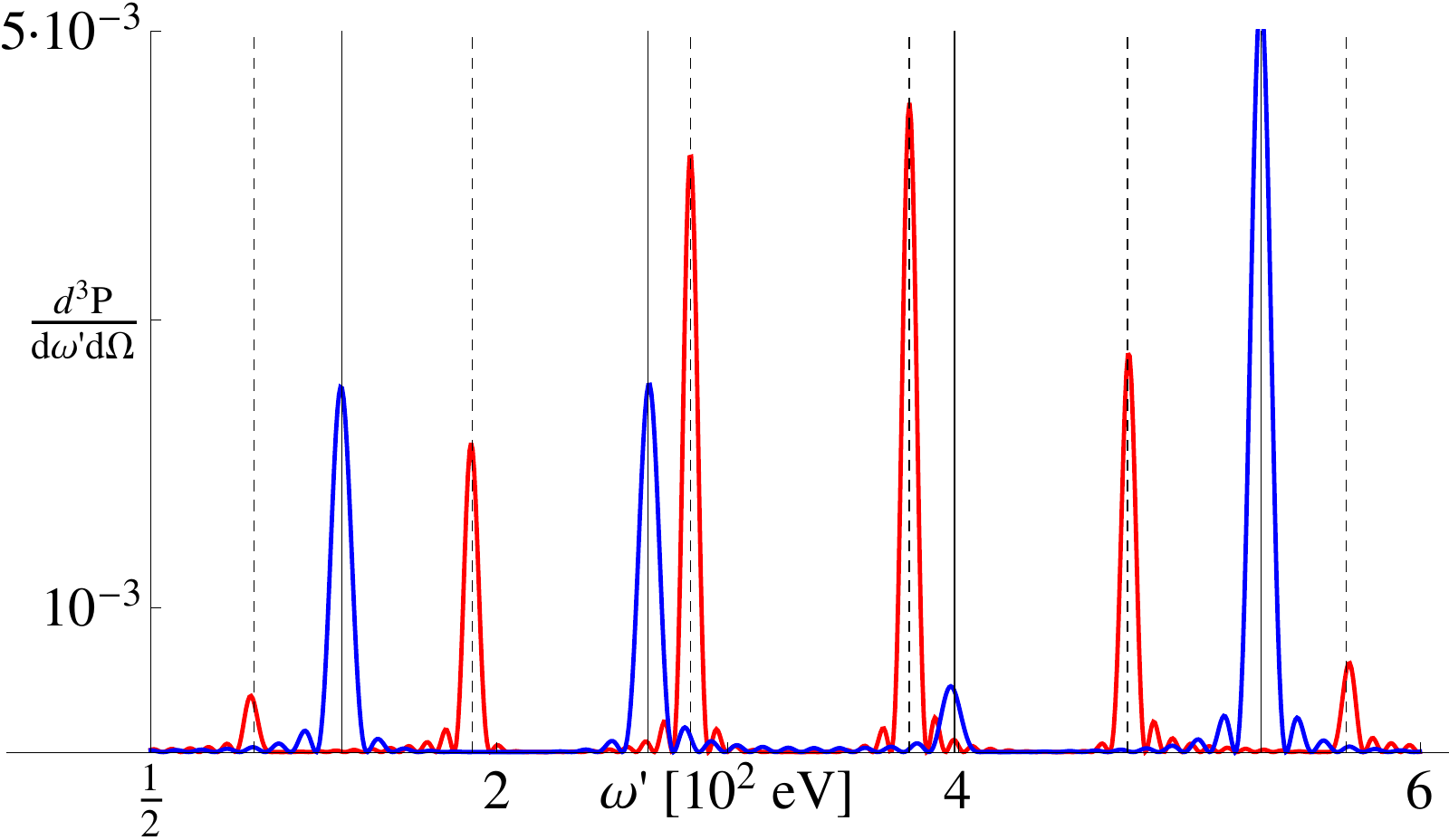}
\caption{\label{FIG:RADIATION} Back-scattered classical spectral density for an electron, $\gamma=10$, colliding at $10^\circ$ to head on with an 800nm laser. We compare $N=10$ cycles of (\ref{PulseZero}) and (\ref{PulseOne}) at $\azero=2$. Radiation angles $\theta=\pi$, $\varphi=0$. Dashed and solid vertical lines show the peaks predicted by the different mass shifts in (\ref{MASSSHIFT}) and (\ref{RED}) respectively.}
\end{figure}
Consider two linearly polarised pulses of equal energy and duration $2\pi N$, both consisting of $N$ cycles of the profiles
\bea
\label{PulseZero}
	f'_1(\phi) &=& \sqrt{2}\sin \phi \;, \\
\label{PulseOne}
%	f'_2(\phi) &=& \sqrt{\tfrac{32}{5}} \sin^2(\tfrac{\phi}{2})\sin(\phi) \;,
	f'_2(\phi) &=& \sqrt{\tfrac{32}{5}} \sin^2(\tfrac{\phi}{2})\sin \phi \;,
\eea
as shown in Fig.~\ref{FIG:ONECYCLE}. The coefficients guarantee that the two pulses contain the same energy, so that we compare like with like.  The first profile is an ordinary sine wave, i.e., a flat-top pulse as discussed above. The second profile is a compressed cycle with a smoother falloff. Writing $f_2'\propto \sin\phi-\tfrac{1}{2}\sin 2\phi$, we see that it describes two {\it co}-propagating waves of different frequencies (a ``two-colour" laser), which has also been explored in the context of pair production  \cite{DiPiazza:2009py}. {Since both these profiles (almost) retain periodicity, we expect both to yield a mass shift, so let us calculate it. Employing (\ref{KINETIC}) we find for the pulse (\ref{PulseZero})},
\be\label{QLINEAR}
	q^1_\mu = p_\mu + \frac{\sqrt{2}\azero m}{k.p}(l.p\, k_\mu - k.p\, l_\mu) + \frac{3\azero^2m^2}{2k.p}k_\mu \;.
\ee
This has both transverse and longitudinal terms (as holds generally; monochromatic fields are a special case \cite{Longer}), but leads to the standard mass shift (\ref{MASSSHIFT}) upon squaring, $q^2=m_*^2$. The quasi-momentum in (\ref{PulseOne}), on the other hand, is
\be\label{QRED}
	q^2_\mu = p_\mu + \frac{3\azero m}{k.p\sqrt{10}}(l.p\, k_\mu - k.p\, l_\mu) + \frac{7\azero^2m^2}{8k.p}k_\mu \;,
\ee
and differs from the monochromatic and flat-top results in both its transverse and longitudinal components. Squaring up, one finds
\be\label{RED}
	%q^2/m^2   =1 + \frac{17}{20}a_0^2 \;,
	q^2 =m^2\big(1 + \tfrac{ 17}{20}\azero^2\big) < m_*^2\;,
\ee
which is a lower mass shift than in both the monochromatic wave and its truncations to finite duration, even though the peak field strength in (\ref{PulseOne}) is higher than in (\ref{PulseZero}). To confirm that the reduced mass shift (\ref{RED}) leads to signals distinct from (\ref{MASSSHIFT}), we consider emission spectra. {The spectral peaks implied by the mass shifts follow from inserting (\ref{QLINEAR}) and (\ref{QRED}) into (\ref{1bad}) and solving for $\omega'$; they are quantitatively different but the explicit expressions for $\omega'_n$ are not too revealing. Instead, we plot an example of the emission rates in Fig.~\ref{FIG:RADIATION}. As we consider only moderate gamma factors with $\gamma\hbar \omega<mc^2$, the spectra are well approximated by the classical limit \cite{Boca:2009zz,Heinzl:2009nd}. We have therefore plotted, for simplicity, the classical emission spectra of a particle in a plane wave, using the textbook methods of \cite[\S 14]{Jackson}. The two different sets of frequencies implied by (\ref{QLINEAR}) and (\ref{QRED}) are marked by vertical lines in the figure, and the peaks in the emission rates are clearly visible at these frequencies.} This shows manifestly that the monochromatic mass shift $m_*$, (\ref{MASSSHIFT}), plays no role in the emission spectrum for the field (\ref{PulseOne}). The new peak pattern cannot be `superimposed' onto that predicted by $m_*$ (for all scattering angles) by a rescaling of the energy in the pulse. Since we compare pulses of equal energy and duration, the origin of our mass shift reduction can only be the shaping of the pulse: it is easily verified that if one replaces $\sin^2$ in (\ref{PulseOne}) with $\sin^{2k}$  (going to a pulse consisting of a train of short, tight peaks) then the mass shift decreases further: for $k=2$, the coefficient of $\azero^2$ is $131/189 < 17/20$.
%
%
%%%%%%%%%%%%%%
\paragraph{General pulses.}
%%%%%%%%%%%%%%
%
%
\begin{figure}[t!]
\includegraphics[width=0.7\columnwidth]{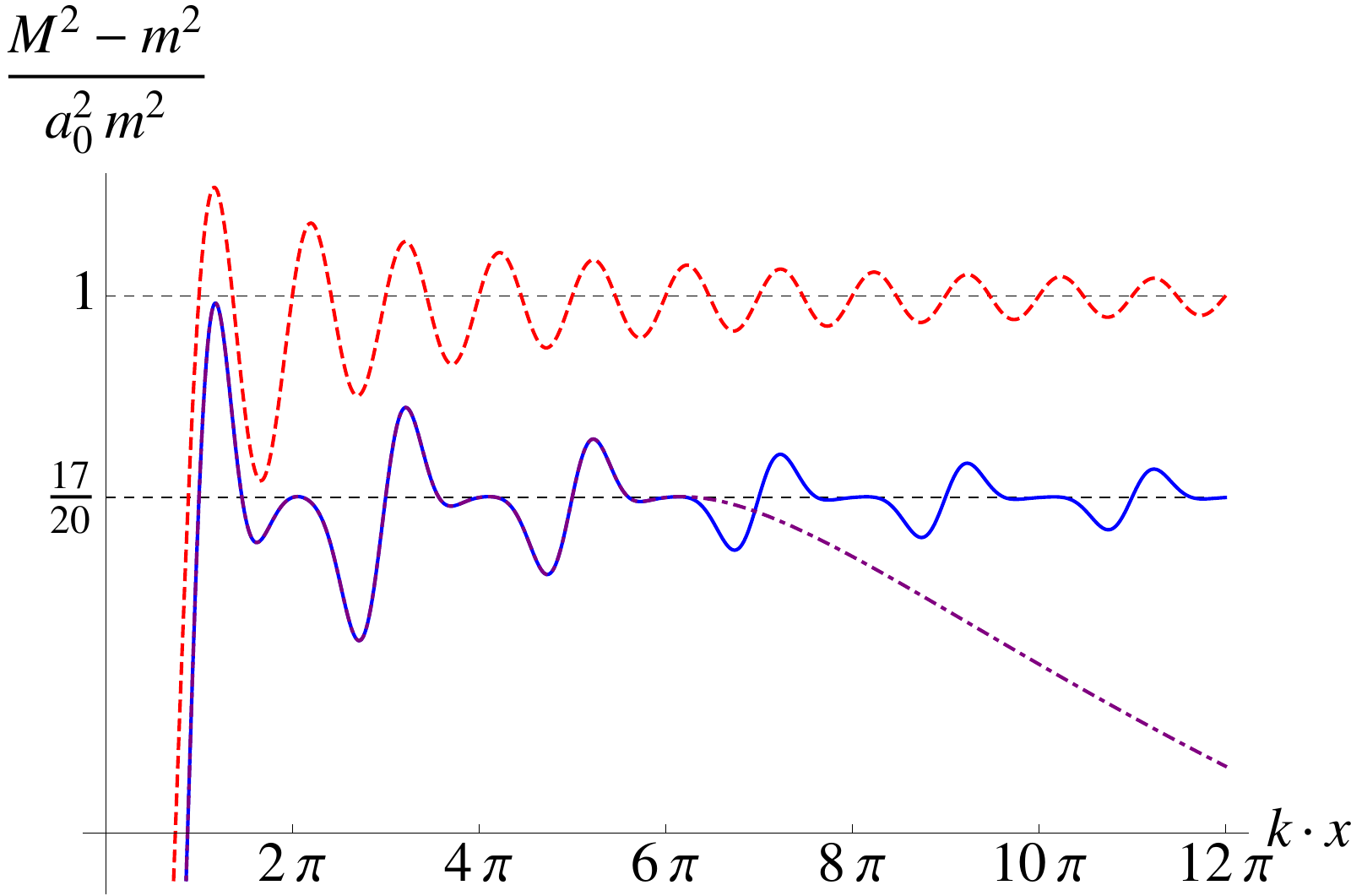}
\caption{\label{FIG:REDUCTION} $M^2(0,k.x)$ for a sinusoidal field [red/dashed], the periodic profile (\ref{PulseOne}) [blue/solid], and three cycles of (\ref{PulseOne}) [purple/dot-dashed]. The monochromatic and reduced mass shifts are shown.}
\end{figure}
{We have now shown that the mass shift $m_*$ is neither unique nor universal. The existence and definition of a mass shift in general pulses may be analysed through the floating average ``$\llangle\ \rrangle$" between arbitrary (lightfront) times $k\!\cdot\! x$ and $k\!\cdot y$ \cite{Brown:1964zz};} the ensuing quasi-momentum squared generalises (\ref{MASSSHIFT}) and (\ref{RED}) to the floating \emph{variance} of the integrated field strength,
\be
	q^2 = M^2(k.x,k.y) \equiv m^2\big(1 + a_0^2 \llangle{f^2}\rrangle - a_0^2\llangle{f}\rrangle\llangle{f} \rrangle\big) \; .
\ee
This $M^2$ appears in the gauge invariant part of the Volkov propagator \cite{Kibble:1975vz} and in the Wigner function \cite{Hebenstreit:2010cc}. We show in Fig.~\ref{FIG:REDUCTION} the function $M^2$ for both (\ref{PulseZero}) and (\ref{PulseOne}). For an infinite number of cycles, $M^2$ increases rapidly from $m^2$ and then oscillates around the appropriate  mass shifts (\ref{MASSSHIFT}) and (\ref{RED}) squared, to which it converges when the time averaged over becomes large.  It is not possible, though, to associate the asymptotic limit of $M$ with a mass shift in general, as this can be shown to be zero for any field of finite (or effectively finite) duration \cite{Hebenstreit:2010cc}. This is also shown in Fig.~\ref{FIG:REDUCTION} for three cycles of (\ref{PulseOne}) as the dot-dashed line: this begins by following the periodic result before falling back to zero. {Rather, it is the approximate plateau in $M^2$, as in Fig.~\ref{FIG:REDUCTION}, which, if prominent enough, implies mass shift signals in the spectrum. This will be investigated in  \cite{Longer}.}
%
%%%%%%%%%%%%%%%
\paragraph{Beam parameters.}
%%%%%%%%%%%%%%%
%
%
Finally, we turn to the parameters required for a measurement of the mass shift.  Consider probe electrons colliding with a laser pulse. In order for the electrons to see only the plane wave (longitudinal) character of the laser, their transverse escape time should be large compared to the time spent in the pulse.  This requires the laser focus to be much wider than the electron beam. The required parameters were previously identified in \cite{Heinzl:2009nd}, and are now realised at experiments such as REGAE at DESY \cite{REGAE}.

\begin{figure}[t!]
\includegraphics[width=\columnwidth]{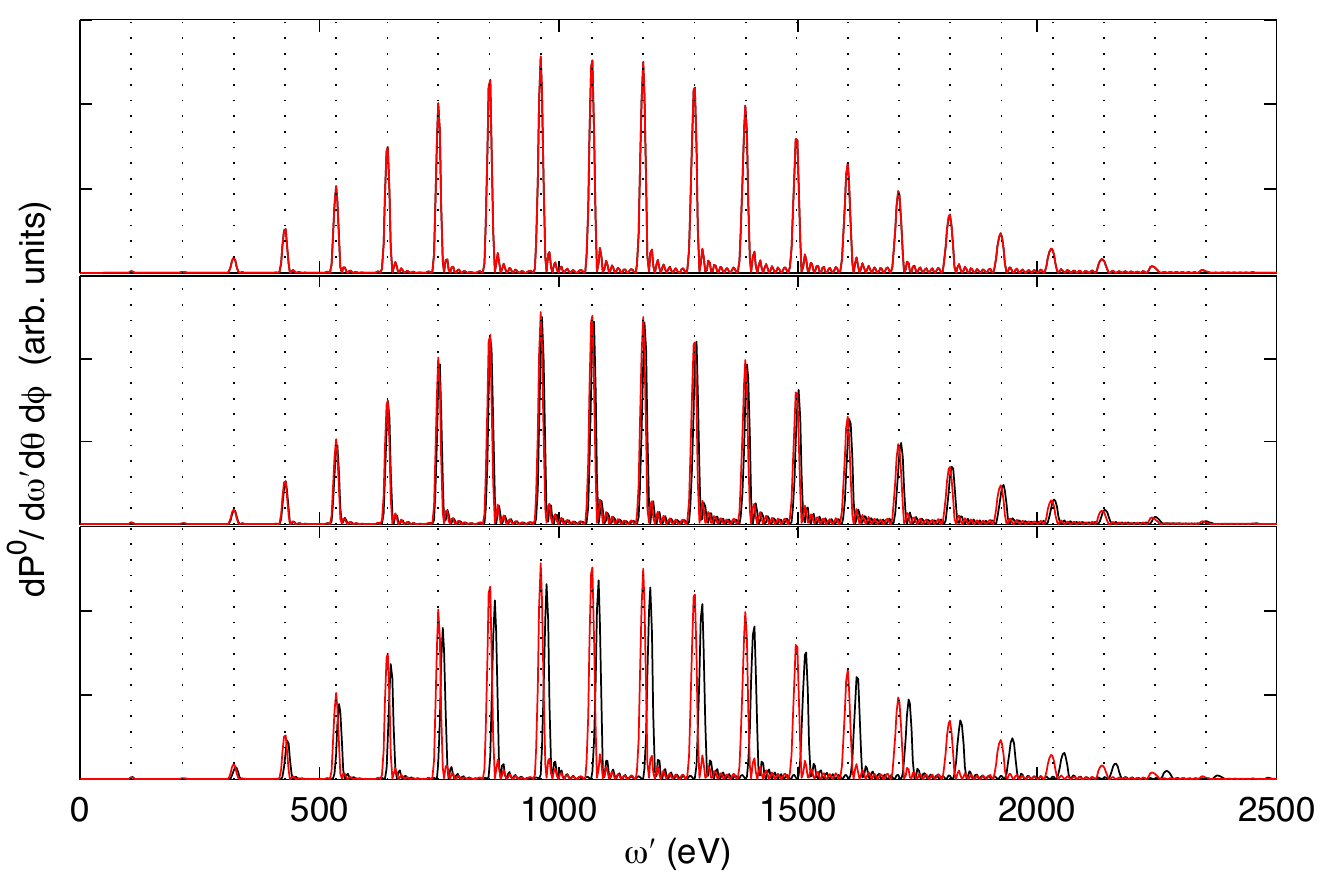}
\caption{\label{GAUSSIAN-PLOTS} Classical spectral density of radiation emitted from electrons in the REGAE setup. The three panels correspond to electrons at 0, 3, 6$\mu m$ from the centre of the REGAE beam. In each panel, the plane wave model is shown in red, the Gaussian beam model in black.}
\end{figure}

The REGAE electron gun can produce a 5 MeV electron beam of width $r_0 \sim 8$ $\mu$m. The laser system is a 200~TW Ti:Sapphire, frequency $\omega=1.55$ eV and focal spot radius $w_0 \sim 40$ $\mu$m. This corresponds to a peak intensity of $ \apeak\sim 2$ \cite{A0}, placing us at the edge of the non-perturbative regime. We consider colliding the laser {(linear polarisation)} and electron beams (at an angle of $10^\circ$ to head on) and measuring properties of the emitted radiation. In Fig.~\ref{GAUSSIAN-PLOTS} we compare the classical spectral density predicted by two models of the laser pulse: a paraxial Gaussian beam and a plane wave, both with the same super-Gaussian (degree 12, $\exp\{-c(\phi)^{12}\}$) profile in the longitudinal (i.e.\ plane wave) direction. The bulk of the electrons in the beam, those with an impact factor below $5\mu$m, are completely blind to the transverse structure of the laser, since $w_0 \gg r_0$. The flat-top section of the super-Gaussian profile contains 10 cycles, giving a duration of around $27$~fs. For a particle entering this section of the pulse the peaks predicted by the flat-top plane wave model and its mass shift are shown by dashed vertical lines in Fig.~{\ref{GAUSSIAN-PLOTS}}, and match the peaks of the spectrum. Hence, mass shift effects are clearly visible for appropriately tuned realistic beams.

%
%
%%%%%%%%%%%%%
\paragraph{{Conclusions.}}
%%%%%%%%%%%%%
%
%
We have shown explicitly that the electron mass shift in a strong laser field can be lowered by pulse shaping. This long known ``intensity-dependent mass'' is therefore also pulse-shape dependent: two pulses with the same energy can have different mass shifts. The mass shift in monochromatic waves is therefore neither unique nor universal.

We have also identified the moderate intensity regime for which transverse size effects become negligible. In this regime, photon emission spectra from laser-particle collisions provide unambiguous mass shift signatures. The experimental setup required is mostly modest: there is no need for ultra-high intensities or ultra-short pulses, the latter since multiple cycles of the beam are required for mass shift signals to become clear. (Previous experiments have been successful in a similar regime \cite{Chen:1998, Bamber:1999zt}.)

Beyond this, the spectra may serve to test the limitations of the plane-wave model. Precision measurements in the above regime (were they to become feasible) could be turned into diagnostic tools for laser pulses at higher intensities. Sufficient knowledge of the spectra would provide a `dictionary' for translating spectral features into properties of the laser pulses (as suggested for carrier phase in \cite{Mackenroth:2010jk}). For example, and as we have seen, the spectral peak positions implied by the mass shift contain information on the shape of the pulse. The mass shift in arbitrary pulses will be addressed in \cite{Longer}. 

A.I.\  thanks C.~Murphy and C.~Palmer for extremely useful discussions. The authors are supported by the European Research Council, contract 204059-QPQV (C.H., A.I.\ and M.M.), and the Swedish Research Council, contracts 2010-3727 (C.H.) and 2011-4221 (A.I.).


\begin{thebibliography}{99}
\bibitem{COMP} {Note, however, the Lorentz invariant characterisation of the two situations is different, since the field invariants are zero in a plane wave and non-zero in an undulator; this difference becomes prominent when the undulator parameter and electron gamma are of the same order.}
\end{thebibliography}
\end{document}